\documentclass[a4paper,11pt]{article} 
\pdfoutput=1
\usepackage{jheppub} 
\usepackage{multirow}

\usepackage{color}

 \usepackage{bm}

\newcommand{\be}{\begin{equation}}
\newcommand{\ee}{\end{equation}}
\newcommand{\bea}{\begin{eqnarray}}
\newcommand{\eea}{\end{eqnarray}}

\title{\boldmath The investigations of the $P$-wave $B_s$ states combining quark model
and lattice QCD in the coupled channel framework}

\author[a]{Zhi Yang,}
\author[b,1]{Guang-Juan Wang,\note{Corresponding author.}}
\author[c,1]{Jia-Jun Wu,}
\author[b,d]{Makoto Oka}
\author[e]{and Shi-Lin Zhu}

\affiliation[a]{School of Physics, University of Electronic Science and
Technology of China, \\Chengdu 610054, China}
\affiliation[b]{Advanced Science Research Center, Japan Atomic Energy
Agency, \\Tokai, Ibaraki, 319-1195, Japan}
\affiliation[c]{School of Physical Sciences, University of Chinese Academy of Sciences (UCAS), \\Beijing 100049, China}
\affiliation[d]{Nishina Center
for Accelerator-Based Science, \\RIKEN, Wako 351-0198, Japan}
\affiliation[e]{School of Physics and Center of High Energy Physics,
Peking University, \\Beijing 100871,China}

\emailAdd{zhiyang@uestc.edu.cn}
\emailAdd{wgj@pku.edu.cn}
\emailAdd{wujiajun@ucas.ac.cn}
\emailAdd{oka@post.j-parc.jp}
\emailAdd{zhusl@pku.edu.cn}

\abstract{Combining the quark model, the quark-pair-creation mechanism and
$B^{(*)}\bar K$ interaction, we have investigated the near-threshold
$P$-wave $B_s$ states in the framework of the Hamiltonian effective
field theory. With the heavy quark flavor symmetry, all the parameters are determined in the $D_s$ sector by fitting the
lattice data.
The masses of the bottom-strange partners of the $D^{*}_{s0}(2317)$
and $D^{*}_{s1}(2460)$ are predicted to be $M_{B^{*}_{s0}}=5730.2_{-1.5}^{+2.4}$ MeV and $M_{B^{*}_{s1}}= 5769.6_{-1.6}^{+2.4}$  MeV, respectively, which are well consistent with the lattice QCD simulation. The two $P$-wave  $B_s$ states
 are the mixtures of the bare $\bar b s$ core and $B^{(*)}\bar K$
component.
Moreover, we find a crossing point between the energy levels with
and without the interaction Hamiltonian in the finite volume
spectrum in the $0^+$ case, which corresponds to a CDD
(Castillejo-Dalitz-Dyson) zero in the $T$-matrix of the $B\bar K$
scattering. This CDD zero will help deepen the insights of the
near-threshold states and can be examined by future lattice
calculation.}

\begin{document} 
\maketitle
\flushbottom

\section{Introduction}
Significant progress has been made in hadron spectroscopy since
2003.
A number of new hadrons involving heavy quarks have been discovered.
However, after 20 years, their properties are still poorly
understood~\cite{ParticleDataGroup:2020ssz}.
Among them, the $D_{s0}^*(2317)$~\cite{BaBar:2003oey} and
$D^*_{s1}(2460)$~\cite{CLEO:2003ggt} are of great interest since
they are much lighter than the quark model
predictions~\cite{Godfrey:1985xj}.
These two states have been widely investigated by both theoretical
and experimental sides, see
reviews~\cite{Chen:2016spr,Dong:2017gaw,Guo:2017jvc,Yao:2020bxx} for
more details.
Various proposals regarding their nature are proposed, including the
quenched and unquenched $c\bar s$ quark
models~\cite{Godfrey:1985xj,Dai:2003yg,Colangelo:2003vg,Hwang:2004cd,Simonov:2004ar,Cheng:2014bca,Song:2015nia,Cheng:2017oqh,Luo:2021dvj,Zhou:2020moj,Alhakami:2016zqx},
the molecule
model~\cite{Kolomeitsev:2003ac,Szczepaniak:2003vy,Hofmann:2003je,vanBeveren:2003kd,
Barnes:2003dj,Gamermann:2006nm,Guo:2006rp,Guo:2006fu,Flynn:2007ki,
Faessler:2007gv,Guo:2009ct,Xie:2010zza,Cleven:2010aw,Wu:2011yb,Guo:2015dha,Albaladejo:2016hae,Du:2017ttu,Guo:2018tjx,Albaladejo:2018mhb,Wu:2019vsy,Kong:2021ohg,Gregory:2021rgy,Wang:2012bu,Huang:2021fdt,Guo:2018tjx},
the tetraquark
model~\cite{Cheng:2003kg,Chen:2004dy,Dmitrasinovic:2005gc,Kim:2005gt,Zhang:2018mnm},
and the $c\bar s$ plus tetraquark
model~\cite{Terasaki:2003qa,Browder:2003fk,Maiani:2004vq,Dai:2006uz,Simonov:2004ar}.

The debates on the inner structures of $D_{s0}^*(2317)$ and
$D^*_{s1}(2460)$ really deepen our understanding on the formation of
a physical state.
At first, the pure quark model gives predictions of four $P$-wave $c\bar s$ mesons with the spin-parity as $J^P=0^+$ ($D^*_{s0}$), $1^+$ ($D^*_{s1}$), $1^+$ ($D^{*'}_{s1}$) and $2^+$ ($D^*_{s2}$). The predicted masses of the higher $D^{*'}_{s1}$ and $D^*_{s2}$ are well consistent with the experimental data, while the two lower ones are not~\cite{ParticleDataGroup:2020ssz, Godfrey:1985xj}. The  $J^P=0^+$ state around 2480
MeV was predicted in the $c\bar{s}$ sector~\cite{Godfrey:1985xj},
which was obviously higher than the state $D^*_{s0}(2317)$
discovered by the experiment later.
A similar situation happened to the $D^*_{s1}(2460)$. What makes
the obscure is that they are very close to the $D^{(*)}K$
threshold.
Using the scattering potential of the light pseudoscalar meson off
the heavy meson based on the Chiral effective field theory
($\chi$-EFT), the $D_{s0}^*(2317)$ and $D^*_{s1}(2460)$ were
described as the dynamically generated bound
states~\cite{Guo:2006fu,Guo:2006rp,Cleven:2010aw}, which indicates
that the interaction at the hadronic level plays an important role
to form these states.
In an alternative view, the predictions of the $c\bar s$ state in
the quark model should provide useful information for the $D_s^*$
states due to its great success in describing the other
conventional hadrons.
However, it should be emphasized that the current quark model is not
perfect for the real world even in a phenomenology study, because
the interaction with the hadron channel is absent there.
Such the hadron channel will play an important role in the
near-threshold state, which is known as the coupled-channel effect.
This effect may shift the masses of the near-threshold hadrons
sizably~\cite{Thomas:1982kv, Thomas:1983fh, Ericson:1983um, Zhu:1998wy, Pennington:2007xr, Zhou:2011sp}.
From this point of view, in a previous paper~\cite{Yang:2021tvc}, we
stuck to the conventional quark model and considered the
coupled-channel effects from the $S$-wave $D^{(*)}K$ channels to
investigate the $D^*_{s0}(2317)$ and $D^*_{s1}(2460)$ states. The higher $D_{s 1}(2536)$ and $D_{s 2}^{*}(2573)$ mainly coupled to the $D$-wave $D^{(*)}K$ channels.
It turns out that the bare states play an extremely important role
in the formation of the physical states.

The extended Hamiltonian effective field theory
(HEFT)~\cite{Hall:2013qba, Wu:2014vma, Hall:2014uca, Liu:2015ktc}
can be used to quantitatively calculate the energy levels and
scattering amplitudes in terms of the hadronic degrees of freedom,
which naturally includes the coupled channel effects of the bare
states and various channels.
In this framework~\cite{Yang:2021tvc}, the bare states and the
mesons in threshold channels are well-defined in the quark model
based on the well-established ground hadron spectrum.
Their coupling potentials can be described by the
quark-pair-creation (QPC) model~\cite{Micu:1968mk} and the
channel-channel interactions can be induced by exchanging the light
mesons.
All the parameters, such as the cutoff and channel coupling
constants will be determined by fitting the lattice simulation
results.
This framework can connect various physical information rather than
just fit the limited data, therefore sufficiently reducing the number
of free parameters.

The bottom analogues of the $D_{s0}^*(2317)$ and $D^*_{s1}(2460)$
are still absent in experiments.
Within the heavy quark symmetry, they are directly related to the
$D_s$ states and the predictions are usually obtained as a by-product in the theoretical study of the $D_s$ states (more
details referred to reviews
\cite{Chen:2016spr,Dong:2017gaw,Guo:2017jvc,Yao:2020bxx,
Meng:2022ozq}).
Thus, the investigations of the bottom analogs can not only enrich
the hadron spectroscopy, but also can be used to examine the
theoretical studies of the near-threshold hadrons.
Nowadays, they have attracted more and more interest.
Their masses have been studied in several scenarios, such as the
$\bar b s$ meson in constitute quark
model~\cite{DiPierro:2001dwf,Ebert:2009ua,Sun:2014wea,li:2021hss},
the $B^{(*)}\bar K$
molecules~\cite{Kolomeitsev:2003ac,Guo:2006fu,Guo:2006rp,Cleven:2010aw,Colangelo:2012xi,Altenbuchinger:2013vwa,Sun:2018zqs},
and the $\bar b s$ plus $B^{(*)}\bar K$
molecules~\cite{Albaladejo:2016ztm}.
In this work, we will use the extended Hamiltonian effective field
theory (HEFT), which has been used in the $D_s$ sector, to study the
bottom analogs.
In our previous work~\cite{Yang:2021tvc}, it was shown that the
heavy quark symmetry is a good symmetry in the $D_s$ sector.
Within the heavy quark symmetry, we will use the same coupling
constants and cutoff parameters in the bottom-strange sector and
obtain the predictions of the spectra of the $P$-wave $B_s$ states.

This paper is arranged as follows.
In Sec.~\ref{sec:HEFT}, we present the HEFT framework in the finite
volume and the $T$-matrix in the infinite volume.
In Sec.~\ref{sec:bare}, we demonstrate the study on the
bottom-strange bare state in the quark model and its coupling with
the nearby threshold channels.
The channel-channel interaction is illustrated in
Sec.~\ref{sec:two}.
In Sec.~\ref{sec:results}, we obtain the mass spectra of the $0^+$
and $1^+$ $B_s$ states, and compare our predicted energy levels with
those from lattice simulation.
At last, a summary is given in Sec.~\ref{sec:summary}.

\section{The Extended Hamiltonian effective field theory} \label{sec:HEFT}

For a physical hadron with multiple components, the
energy-independent Hamiltonian reads
\begin{equation} \label{eq:hamiltonian}
H=H_0+H_I\,,
\end{equation}
where $H_0$ is the non-interacting Hamiltonian,
\begin{equation}\label{eq:qmhamiltonian}
H_0=\sum_{b} |b\rangle\, m_{b}\,\langle b | + \sum_\alpha \int d^3
\vec{k} \,|\alpha(\vec{k})\rangle \,E_\alpha(\vec{k})\, \langle
\alpha(\vec{k})|\,.
\end{equation}
Here $b$ represents a bare $\bar b s$ core with a mass $m_{b}$, which
is defined in the quark model.
$\alpha$ denotes the $B^{(*)}\bar K$ channels, and
$E_\alpha(\vec{k})=\sqrt{m_K^2+\vec{k}^2}+\sqrt{m^2_{B^{(*)}}+\vec{k}^2}$
is the kinematic energy with $\vec{k}$ the relative momentum.
The interacting Hamiltonian is $H_I=g+v$, where $g$ and $v$ are the
potential between the bare $\bar b s$ core and $B^{(*)}\bar K$ channels,
and the potential within the  $B^{(*)}\bar K$ channels, respectively.
Their explicit forms will be illustrated in the following sections.

\subsection{The bare state $\bar b s$}\label{sec:bare}

Since the quark model gains a great success in explaining the
properties of the low-lying mesons, the predicted  $\bar b s$ state
is naturally expected to exist.
In this work, we adopt the relativized quark model proposed by
Godfrey-Isgur (GI)~\cite{Godfrey:1985xj} to determine the $\bar b s$
state.
The GI model provided a successful description of the mass spectra
of the low-lying mesons, from the pion to the bottomonium
\cite{Godfrey:1985xj}.

In the quark model, the Hamiltonian describing quark-antiquark
interaction reads,
\begin{equation}
 \mathcal{H}_{q\bar q}=\sqrt{{\bf p}^2+m_q^2}+\sqrt{{\bf p}^2+m_{\bar q}^2}+\frac{{\bf \lambda}_q}{2}\frac{{\bf \lambda}_{\bar q}}{2}\mathcal{V}_{q\bar q},
 \label{eq:schrodinger}
\end{equation}
where  the ${\bf \lambda}_{q(\bar q)}$ and $m_{q(\bar q)}$  are the
color matrix and the mass of the constituent quark (anti-quark),
respectively.
The effective potential $\mathcal{V}_{q \bar q}$ contains the
one-gluon-exchange interaction and linear confinement interaction.
Its explicit form can be found in Ref.~\cite{Godfrey:1985xj}.

\begin{table}
 \renewcommand\arraystretch{1.5}
\label{tab:parameters}
 \centering
  \setlength{\tabcolsep}{5.5mm}
\begin{tabular}{cccc}
\hline\hline
& parameter & this work &  GI~\cite{Godfrey:1985xj} \\ \hline\hline
Masses & $\frac{1}{2}\left(m_{u}+m_{d}\right)$ & 264 MeV & 220 MeV \\
 & $m_{s}$ & 497 MeV  & 419 MeV \\
 & $m_{c}$ & 1720 MeV & 1628 MeV \\
 & $m_{b}$ & 5065 MeV & 4977 MeV \\
 \hline
Potentials & $b$ & 0.18 $\mathrm{GeV}^{2}$ & 0.18 $\mathrm{GeV}^{2}$ \\
 & $c$ & $-426$ MeV & $-253$ MeV \\
\hline
Relativistic effects & $\sigma_{0}$ & 1.45 GeV & 1.80 GeV \\
 & $s$ & 1.55 & 1.55 \\
 & $\epsilon_{c}$ & $-0.194$ & $-0.168$ \\
 & $\epsilon_{t}$ & $-0.016$ & 0.025 \\
 & $\epsilon_{\text {so}(V)}$ & $-0.277$ & $-0.035$ \\
 & $\epsilon_{\text {so }(S)}$ &$ -0.289$ & 0.055 \\
\hline\hline
\end{tabular}
\caption{The free parameters in the potential quark model.}
\end{table}

\begin{figure}[t]
\centering
\includegraphics[width=0.6\linewidth]{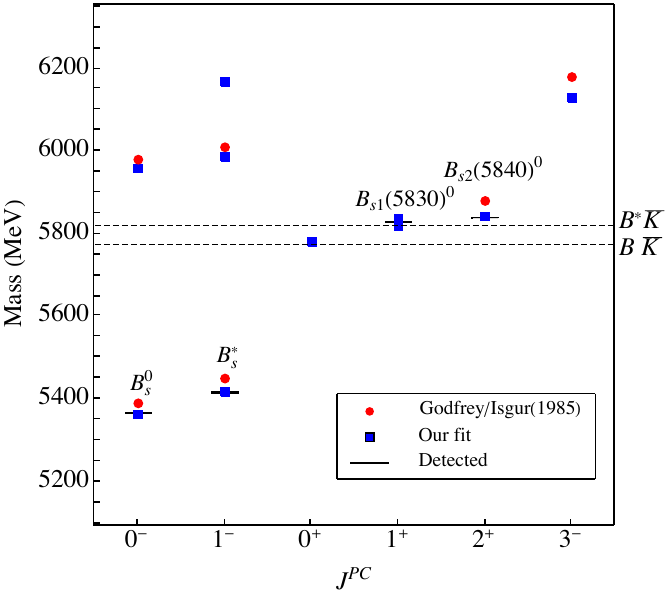}
\caption{
Mass spectrum of bare $\bar b s$ mesons within the relativized quark
model.
The circles and squares are the results predicted in
Ref.~\mbox{\cite{Godfrey:1985xj} }\hspace{0pt} and our new fit,
respectively.
The shaded areas represent the experimental masses and their
uncertainties~\cite{ParticleDataGroup:2020ssz}.} \label{fig:Bs}
\end{figure}

In Ref.~\cite{Yang:2021tvc}, we used the masses of the
well-established mesons which locate far away from the two-meson
thresholds to update the parameters in the GI model.
Their values are summarized in Table~\ref{tab:parameters}.
The mass spectrum is better fitted to the experimental data than
that in Ref.~\cite{Godfrey:1985xj}.
In Fig.~\ref{fig:Bs}, we present the comparison of the mass spectrum
for the bottom-strange mesons using the original and updated set of
parameters.
In the quark model, there are four $P$-wave $\bar b s$ mesons. 
These bare $\bar b s$ states are in the vicinity of
the $B^{(*)}\bar K$ channels, which resembles the charm-strange case.
Thus, the coupled-channel effects may significantly shift the masses
of these $B_s$ states.
The possible coupling channels are shown in Table~\ref{tab:baremass}.
Similar to the $D_s$ states, the lighter and heavier bare $1^+$
states are almost on the heavy quark spin bases, which implies a
good heavy quark symmetry.
Thus, the bare $0^+$ and the lighter bare $1^+$ states are expected to mainly couple with the $S$-wave $B\bar K$ and $B^*\bar K$ channels, respectively, while the heavier $1^+$ and the $2^+$ bare states mainly couple with the $D$-wave $B^*\bar K$ or $B\bar K$ channels within good heavy quark symmetry.  
The $D$-wave coupling is significantly suppressed in the vicinity of the thresholds.
Therefore, the mass shifts of the heavier bare $1^+$ and $2^+$ states can be neglected. 
Indeed, their bare masses, $5835.6$ MeV and $5842.7$ MeV, are very close to the experimental ones $5828.70 \pm 0.20$ MeV and $5839.86 \pm 0.12$ MeV \cite{ParticleDataGroup:2020ssz}.

\begin{table}
 \renewcommand\arraystretch{1.5}
\label{tab:baremass}
 \centering
\begin{tabular}{rccccc}
\hline\hline
&  &  $b\,(|^{2S+1}L_J\rangle)$                        & $b\,$(mass) & $\alpha$ & $L$ \\
\hline
 &  $B_{s0}^*$   & $|^3P_0\rangle$                              & 5780.9 & $B\bar K$ & $S$  \\
 &  $B_{s1}^*$ & $-0.74\,|^1P_1\rangle+0.67\,|^3P_1\rangle$    & 5818.5  & $B^*\bar K$ & $S,\,D$ \\
       & & $=0.98\,\phi_s\, - 0.22\,\phi_d$  &  &      &\\
$   $    &  $B_{s1}^{*\prime}$ & $0.67\,|^1P_1\rangle+0.74\,|^3P_1\rangle$    & 5835.6 & $B^*\bar K$  & $S,\,D$ \\
       && $=0.22\,\phi_s\, + 0.98\,\phi_d$  &  &      &\\
       
   $   $    &  $B_{s2}^{*\prime}$  & $|^3P_2\rangle$            & $5842.7$  & $B\bar K$, $B^*\bar K$  & $D$ \\
\hline\hline
\end{tabular}
\caption{
The related bare $\bar b s$ cores ($b$)  and the $B^{(*)}\bar K$
($\alpha$) channels in the Hamiltonians of the physical $B_s$
states.
The wave functions and mass spectrum (MeV) of the bare states are
shown.
$\phi_s=|\frac{1}{2}_l\otimes \frac{1}{2}_h\rangle$ and
$\phi_d=|\frac{3}{2}_l\otimes \frac{1}{2}_h\rangle$ are  the heavy
quark symmetry bases, where $h$ and $l$ are the heavy and light
degrees of freedom, respectively.
The script $L$ in the last column denotes the orbital excitation in
the $B^{(*)}\bar K$ channels. }
\end{table}

\begin{figure}[t]
\centering
\includegraphics[width=0.45\linewidth]{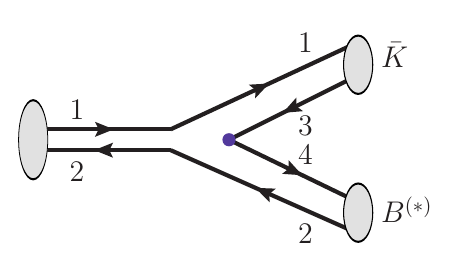}
\caption{ The diagram contributes to the process of the bare state coupling to $B^{(*)}\bar K$ channel in the QPC model. The first and second quarks are anti-strange and bottom quarks, respectively. The third and fourth are the light quarks created from a vacuum. }
\label{fig:3P0}
\end{figure}

In this work, the interactions between $\bar b s$ and $B^{(*)}\bar K$ can be written as
 \begin{eqnarray}
&&g=\sum_{\alpha, b}\int d^3\vec{k} \left\{|\alpha(\vec{k})\rangle
\,g_{\alpha\,b}(\vec{k})\,\langle b |+h.c. \right\}\,,
\end{eqnarray}
where $g_{\alpha\,b}(\vec{k})$ can be calculated from its partial wave expansion $g^{L_\alpha}_{\alpha\,b}(|\vec{k}|)$.
The $\bar b s$ state couples with the $B^{(*)}\bar K$ channels through
the creation of a light quark-antiquark pair with a quantum number
$J^{PC}=0^{++}$, as shown in Fig.~\ref{fig:3P0}.
From the phenomenological QPC model
\cite{LeYaouanc:1977fsz,Kokoski:1985is,Page:1995rh,Blundell:1996as,Ackleh:1996yt,Morel:2002vk,Ortega:2016mms}, $g^{L_\alpha}_{\alpha\,b}(|\vec{k}|)$ reads
 \begin{equation}
g^{L_\alpha}_{\alpha\,b}(|\vec{k}|)=\gamma I^{L_\alpha}_{\alpha\,b}(|\vec{k}|)
e^{-\frac{\vec{k}^2}{2\Lambda^{\prime 2}}}.
\end{equation}
Here, $\gamma$ is related to the creation probability of the quark
pair.
The exponential form factor with the cutoff $\Lambda^{\prime}$ is
introduced to truncate the hard vertices~\cite{Morel:2002vk,
Ortega:2016mms}.
The spatial transform factor $I^{L_\alpha}_{\alpha\,b}(|\vec{k}|)$ can be
calculated with the exact wave functions obtained from the quark
model as
\begin{eqnarray}
I_{\alpha\,b}^{L_{\alpha}}(|\vec{k}|)&=&-\frac{\sqrt{4\pi(2L_{\alpha}+1)}}{2J_{b}+1}\sum_{M_{J_{\alpha_{1}}}M_{J_{\alpha_{2}}}}C_{L_\alpha0;J_{\alpha_{12}}(M_{J_{\alpha_{1}}}+M_{J_{\alpha_{2}}})}^{J_{b}(M_{J_{\alpha_{1}}}+M_{J_{\alpha_{2}}})}C_{J_{\alpha_{1}}M_{J_{\alpha_{1}}};J_{\alpha_{2}}M_{J_{\alpha_{2}}}}^{J_{\alpha_{12}}(M_{J_{\alpha_{1}}}+M_{J_{\alpha_{2}}})}\nonumber \\
&\times&\sum \mathcal{A}(M_{J_{\alpha_{1}}},M_{J_{\alpha_{2}}},M_{S_{b}},M_{L_{b}},M_{L_{\alpha_{1}}},M_{S_{\alpha_{1}}},M_{L_{\alpha_{2}}},M_{S_{\alpha_{2}}},M)
\nonumber\\
&\times&\int\mathrm{d}^{3}p\;\psi_{n_{\alpha_{1}}L_{\alpha_{1}}M_{L_{\alpha_{1}}}}^{*}\left(\frac{m_{3}}{m_{2}+m_{3}}\vec{k}+\vec{p}\right)\psi_{n_{\alpha_{2}}L_{\alpha_{2}}M_{L_{\alpha_{2}}}}^{*}\left(\frac{m_{3}}{m_{1}+m_{3}}\vec{k}+\vec{p}\right)\nonumber 
\\
 & \times&\psi_{n_{b}L_{b}(M_{L_{\alpha_1}}+M_{L_{\alpha_2}})}(\vec{k}+\vec{p})\mathcal{Y}_{LM}(\vec{p}), 
 \end{eqnarray}
where the $S_b$$ (M_{S_b})$, $L_b(M_{L_b})$, and $J_b$ are the intrinsic spin, the orbital angular momentum (third direction component) and the total spin of the $b$-th bare core, respectively.  $S_{\alpha_{1/2}}$ ($M_{S_{\alpha_{1/2}}}$),  $L_{\alpha_{1/2}}$ ($M_{L_{\alpha_{1/2}}}$), and $J_{\alpha_{1/2}}$ ($M_{J_{\alpha_{1/2}}}$) are corresponding quantum numbers of the two mesons in the $\alpha$-th $B^{(*)}\bar K$ channels, respectively.  $C$ is the 3-j Clebsch–Gordan coefficients.  The two Clebsch–Gordan coefficients show the couplings   $\vec{J}_{\alpha_{12}}=\vec{J}_{\alpha_1}+\vec{J}_{\alpha_2}$  and $\vec{J}_{b}=\vec{J}_{{\alpha_{12}}}+\vec{L}_{{\alpha}}$ (  $L_{\alpha}$ being the orbital excitation between two mesons in the $\alpha$-th channel). $L(M)$ denotes the relative orbital angular momentum between the third and the fourth quarks created in the vacuum as illustrated in Fig. \ref{fig:3P0}. The coefficient $\mathcal A$ denotes a series of Clebsch–Gordan coefficients for the coupling of the spin and orbital angular momenta (details are referred to Refs. \cite{osti_7023003, Roberts:1992esl}). In the spatial integrals, the $\mathcal{Y}_{LM}(\vec{p})=|\vec{p}|^L Y_{L M}(\hat{p})$ ($p$ is the relative momentum) is the relative wave function between the third and the fourth quarks created in the vacuum. $Y_{LM}$ is the spherical harmonics wave function.  The $\psi_{nlm}$ is the spatial wave function of the meson and reads~\cite{Wong:2001td}
\begin{equation}
\psi_{nlm}(\vec{p})=N_{n}|\vec{p}|^l \sqrt{\frac{4 \pi}{(2 l+1) ! !}} Y_{lm}(\hat{p}) \exp \left\{-\frac{\vec{p}^{\;2}}{2 n \beta^{2}}\right\}, \nonumber
\end{equation}
where $\beta$ is the oscillating parameters, and $n$ is related to the radial excitation. The normalization  factor is 
\begin{equation}
N_{n}=\left(\frac{1}{\pi n \beta^{2}}\right)^{3 / 4} \frac{1}{\left(2 n \beta^{2}\right)^{l / 2}}. \nonumber
\end{equation}

The $\bar b s $ may also couple with the other channels.
However, the couplings $\bar b s\rightarrow B^{*}_s\pi$ or $\bar b s\rightarrow B_s \gamma$ can be neglected, since the strengths of
the isospin-breaking and electromagnetic vertices are significantly
weaker than the strong one.
Other possible strongly coupled channels, such as the $B_s\eta$ for
$B^*_{s0}$, are located far away from the physical states and
therefore are not considered.

\subsection{Two-body potential} \label{sec:two}

The potential within the two-body channels reads
\begin{eqnarray}
v=\sum_{\alpha,\,\beta}\int d^3\vec{k}\,d^3\vec{k}'
\,|\alpha(\vec{k})\rangle\,
V_{\alpha,\,\beta}(\vec{k},\,\vec{k}')\,\langle
\beta(\vec{k}')|\,,
\end{eqnarray}
where $V_{\alpha,\,\beta}(\vec{k},\,\vec{k}')$ can be straightforwardly obtained by the
Lagrangian~\cite{Lin:1999ad, Oset:2010tof, Zhao:2014gqa},
\begin{eqnarray}
\mathcal L &=& \mathcal L_{PPV}+\mathcal L_{VVV}\nonumber\\
&=&i g_v \text{Tr} (\partial^\mu P\,[P,V_\mu]\,)+ig_v\text{Tr}
(\partial^\mu V^\nu\,[V_\mu,V_\nu]\,)\,, \label{eq:lag}
\end{eqnarray}
where $g_v$ is an overall coupling constant.
In the $SU(4)$ flavor symmetry~\cite{Lin:1999ve}, the $P$ and $V$,
respectively, represent the $4\times 4$ pseudoscalar and vector
meson matrices:
\begin{eqnarray}
P&=&\frac{1}{\sqrt{2}}\left(\begin{array}{cccc}
\frac{\pi^{0}}{\sqrt{2}}+\frac{\eta}{\sqrt{6}}+\frac{\eta_{c}}{\sqrt{12}} & \pi^{+} & K^{+} & \overline{D^{0}} \\
\pi^{-} & -\frac{\pi^{0}}{\sqrt{2}}+\frac{\eta}{\sqrt{6}}+\frac{\eta_{c}}{\sqrt{12}} & K^{0} & D^{-} \\
K^{-} & \bar{K}^{0} & -\sqrt{\frac{2}{3}} \eta+\frac{\eta_{c}}{\sqrt{12}} & D_{s}^{-} \\
D^{0} & D^{+} & D_{s}^{+} & -\frac{3 \eta_{c}}{\sqrt{12}}\end{array}\right)\,,\\
V&=&\frac{1}{\sqrt{2}}\left(\begin{array}{cccc}
\frac{\rho^{0}}{\sqrt{2}}+\frac{\omega^{\prime}}{\sqrt{6}}+\frac{J / \psi}{\sqrt{12}} & \rho^{+} & K^{*+} & D^{* 0} \\
\rho^{-} & -\frac{\rho^{0}}{\sqrt{2}}+\frac{\omega^{\prime}}{\sqrt{6}}+\frac{J / \psi}{\sqrt{12}} & K^{* 0} & D^{*-} \\
K^{*-} & K^{* 0} & -\sqrt{\frac{2}{3}} \omega^{\prime}+\frac{J / \psi}{\sqrt{12}} & D_{s}^{*-} \\
D^{* 0} & D^{*+} & D_{s}^{*+} & -\frac{3 J / \psi}{\sqrt{12}}
\end{array}\right)\,.
 \end{eqnarray}
Considering the sizable $SU(4)$ symmetry breaking, we do not use the
same coupling constant for the $D^{(*)}D^{(*)}V$ and the $KKV$
vertices in the $D^{(*)}K\rightarrow D^{(*)}K$ progresses.
The related coupling constant $g_{c}$ $\left(\propto
g_{D^{(*)}D^{(*)}V}g_{KKV}\right)$ was determined by fitting data in
the $D_s$ sectors, which will be illustrated in
Sec.~\ref{sec:results}.
For the interactions in the $B^{(*)}\bar K\rightarrow B^{(*)}\bar K$
scattering here, we use the same formula as those in the charmed
strange sector, since the heavy $b$ and $c$ quarks are both nice
spectators.
To include the effects of the hadron structures, we
introduce a form factor with a cutoff parameter $\Lambda$ for the
interaction vertex,
\begin{equation}
\left(\frac{\Lambda^{2}}{\Lambda^{2}+p_{f}^{2}}\right)^{2}\left(\frac{\Lambda^{2}}{\Lambda^{2}+p_{i}^{2}}\right)^{2},\label{eq:ff}
\end{equation}
where $p_i$ and $p_f$ are the relative momenta of initial and final particles in the $B^{(*)}K \to B^{(*)}K$ process, respectively.

In the calculation, we need the partial wave expansion of $V_{\alpha,\,\beta}(\vec{k},\,\vec{k}')$, i.e., $V^{JL_\alpha L_\beta}_{\alpha,\,\beta}(|\vec{k}|,\,|\vec{k}'|)$. Here $J$ is the total angular momentum, which equals the spin of the corresponding bare state. $L_\alpha$ and $L_\beta$ are the partial wave quantum numbers of $\alpha$ and $\beta$ channels, respectively.

\subsection{The Hamiltonian in the finite volume}

In a box with length $L$, the possible momentum values are the
integral multiples of the lowest non-trivial momentum $2\pi/L$ in any
one dimension, i.e. $k_{n}=\sqrt{n} \frac{2 \pi}{L},\;
\sqrt{n}=\sqrt{n_x^2+n_y^2+n_z^2},\; n_{x,y,z} = 0,\pm1,\pm2,...$.  The
potentials are then transformed into the discretized forms
\begin{eqnarray}\label{eq:dispotential}
\begin{aligned}
&\tilde{g}^{L_\alpha}_{\alpha\,b}\left(k_{n}\right)=\sqrt{\frac{C_{3}(n)}{4 \pi}}\left(\frac{2 \pi}{L}\right)^{3 / 2} g^{L_\alpha}_{\alpha\,b}\left(k_{n}\right),\\
&\tilde{v}_{L_\alpha\,L_\beta}^{J,\,\alpha,\,\beta}\left(k_{n_{1}},
k_{n_{2}}\right)=\sqrt{\frac{C_{3}\left(n_{1}\right)}{4 \pi}}
\sqrt{\frac{C_{3}\left(n_{2}\right)}{4 \pi}}\left(\frac{2
\pi}{L}\right)^{3} V^{J\,L_\alpha\,L_\beta}_{\alpha,\,\beta}\left(k_{n_{1}}, k_{n_{2}}\right),
\end{aligned}
\end{eqnarray}
where a factor $C_3(n)$ is introduced and it denotes the number of  choices of $n_x, n_y, n_z$ to form $n$.

With the kinematic energy and the discretized potentials, the finite-volume Hamiltonian matrix can be easily obtained.   %
\\
\\
{\bf $0^+$: one bare state and one channel}\\
In this case, the Hamiltonian contains the $\bar b s$ core ($^3P_0$)
and the $B\bar K$ channel as well as their interactions. For simplicity, we denote the potential $\tilde{g}=\tilde{g}^{S}_{B\bar K\,^3P_0}$
and $\tilde{v}=\tilde{v}_{S\,S}^{0,\,B\bar K\,,B\bar K}$.
Solving the Schr\"odinger equation $H|\Psi_E\rangle=E|\Psi_E\rangle$
is equivalent to finding the solutions of the following matrix
equations
\begin{equation}
\operatorname{det}\left(\left[H_{0}\right]_{N+1}+\left[H_{I}\right]_{N+1}-E[I]_{N+1}\right)=0\;,
\end{equation}
where $\operatorname{ det}$ represents taking the determinant of the
matrix and $[I]_{N+1}$ is an $(N+1)\times(N+1)$ unit matrix.
$N$ must be large enough until the results are stable, here we use $N = 600$.
The non-interacting and interacting Hamiltonian matrices can be
written as
\begin{eqnarray}
H_{0}=\left(\begin{array}{cccc}
m_{b} & 0 & 0 & \cdots \\
0 & \sqrt{k_{0}^{2}+m_{B}^{2}}+\sqrt{k_{0}^{2}+m_{K}^{2}} & 0 & \cdots \\
0 & 0 & \sqrt{k_{1}^{2}+m_{B}^{2}}+\sqrt{k_{1}^{2}+m_{K}^{2}} & \cdots \\
\vdots & \vdots & \vdots & \ddots
\end{array}\right)
\end{eqnarray}
and
\begin{eqnarray}
H_{I}=\left(\begin{array}{cccc}
0 & \tilde{g}\left(k_{0}\right) & \tilde{g}\left(k_{1}\right) & \cdots \\
\tilde{g}\left(k_{0}\right) & \tilde{v}\left(k_{0}, k_{0}\right) & \tilde{v}\left(k_{0}, k_{1}\right) & \cdots \\
\tilde{g}\left(k_{1}\right) & \tilde{v}\left(k_{1}, k_{0}\right) & \tilde{v}\left(k_{1}, k_{1}\right) & \cdots \\
\vdots & \vdots & \vdots & \ddots
\end{array}\right)\;,
\end{eqnarray}
respectively.
\\
\\
{\bf $1^+$: two bare states and one channel} \\
There are two bare cores with the spin-parity $J^P=1^+$ in the quark
model, i.e., $^1P_1$ and $^3P_1$.
They will couple with the $S$-wave and $D$-wave $B^*\bar K$ channels.
For simplicity, we denote the potential $\tilde{g}_{i L^\prime}=\tilde{g}^{L^\prime}_{B^*\bar K\,i}$
and $\tilde{v}_{L^\prime\,L^{\prime\prime}}=\tilde{v}_{L^\prime\,L^{\prime\prime}}^{1,\,B^*\bar K\,,B^*\bar K}$. Here $i (=1,2)$ denotes the bare $^1P_1$ and $^3P_1$ states. $L^\prime$ and $L^{\prime\prime}$ could be $S$ or $D$.
This leads to a $(2N+2)\times(2N+2)$ Hamiltonian matrix
\begin{equation}
\operatorname{det}\left(\left[H_{0}\right]_{2N+2}+\left[H_{I}\right]_{2N+2}-E[I]_{2N+2}\right)=0,
\end{equation}
with the explicit forms as
\begin{eqnarray}
H_{0}=\left(\begin{array}{ccccccc}
m_{b1} & 0 & 0 & 0 & 0 & 0 & \cdots \\
0 & m_{b2}  & 0 & 0 & 0 & 0 & \cdots \\
0 & 0 &  E(0)  & 0  & 0 & 0 & \cdots \\
0 & 0 & 0 & E(0) & 0 & 0 & \cdots \\
0 & 0 & 0 & 0 & E(1) & 0 & \cdots \\
0 & 0 & 0 & 0 & 0 & E(1) & \cdots \\
\vdots & \vdots & \vdots & \vdots &\vdots &\vdots & \ddots
\end{array}\right),
\end{eqnarray}
where $E(n)=\sqrt{k_{n}^{2}+m_{B^*}^{2}}+\sqrt{k_{n}^{2}+m_{K}^{2}}$, and
\begin{eqnarray}
H_{I}=\left(\begin{array}{ccccccc}
0 & 0 & \tilde{g}_{1S}\left(k_{0}\right) & \tilde{g}_{1D}\left(k_{0}\right) & \tilde{g}_{1S}\left(k_{1}\right) & \tilde{g}_{1D}\left(k_{1}\right) & \cdots \\
0 & 0 & \tilde{g}_{2S}\left(k_{0}\right) & \tilde{g}_{2D}\left(k_{0}\right) & \tilde{g}_{2S}\left(k_{1}\right) & \tilde{g}_{2D}\left(k_{1}\right) & \cdots \\
\tilde{g}_{1S}\left(k_{0}\right) & \tilde{g}_{2S}\left(k_{0}\right) & \tilde{v}_{SS}\left(k_{0}, k_{0}\right) & \tilde{v}_{SD}\left(k_{0}, k_{0}\right) & \tilde{v}_{SS}\left(k_{0}, k_{1}\right) & \tilde{v}_{SD}\left(k_{0}, k_{1}\right) & \cdots \\
\tilde{g}_{1D}\left(k_{0}\right) & \tilde{g}_{2D}\left(k_{0}\right) & \tilde{v}_{DS}\left(k_{0}, k_{0}\right) & \tilde{v}_{DD}\left(k_{0}, k_{0}\right) & \tilde{v}_{DS}\left(k_{0}, k_{1}\right) & \tilde{v}_{DD}\left(k_{0}, k_{1}\right) & \cdots \\
\tilde{g}_{1S}\left(k_{1}\right) & \tilde{g}_{2S}\left(k_{1}\right) & \tilde{v}_{SS}\left(k_{1}, k_{0}\right) & \tilde{v}_{SD}\left(k_{1}, k_{0}\right) & \tilde{v}_{SS}\left(k_{1}, k_{1}\right) & \tilde{v}_{SD}\left(k_{1}, k_{1}\right) & \cdots \\
\tilde{g}_{1D}\left(k_{1}\right) & \tilde{g}_{2D}\left(k_{1}\right) & \tilde{v}_{DS}\left(k_{1}, k_{0}\right) & \tilde{v}_{DD}\left(k_{1}, k_{0}\right) & \tilde{v}_{DS}\left(k_{1}, k_{1}\right) & \tilde{v}_{DD}\left(k_{1}, k_{1}\right) & \cdots \\
\vdots & \vdots & \vdots &\vdots & \vdots & \vdots & \ddots
\end{array}\right)\;.
\end{eqnarray}
\\
\\
%
The energy levels in the finite volume correspond to the eigenvalues
of the Hamiltonian matrix, which can be used to determine the
parameters by fitting the lattice data.
For the eigenvectors of the Hamiltonian matrix, the squares of its
coefficients represent the probabilities $P(\alpha)$ ($\alpha =\bar b s, B^{(*)}\bar K$) of the bare $b\bar{s}$ and $B^{(*)}\bar K$
components~\cite{Wu:2014vma}.

With the parameters determined, we will return to the infinite limit
to study the physical properties of the scattering $T$-matrix by
solving the relativistic Lippmann-Schwinger
equation~\cite{Matsuyama:2006rp,Wu:2012md,Wu:2014vma,Liu:2015ktc},
\begin{eqnarray}
T^{J\,L_\alpha\,L_\beta}_{\alpha,\,\beta}(k,k';E)&=&{\mathcal
V}^{J\,L_\alpha\,L_\beta}_{\alpha,\,\beta}(k,k';E)+\sum_{\alpha'\,L_{\alpha'}}\int q^2dq
\nonumber\\
&&\times {\mathcal
V}^{J\,L_\alpha\,L_{\alpha'}}_{\alpha,\,\alpha'}(k,q;E)\frac{1}{E-E_{\alpha'}(q)+i\epsilon}T^{J\,L_{\alpha'}\,L_\beta}_{{\alpha'},\,\beta}(q,k';E)\,,
\end{eqnarray}
where the effective potential ${\mathcal
V}_{\alpha,\,\beta}(k,k';E)$ is related to the interaction
Hamiltonian,
\begin{equation}
{\mathcal
V}^{J\,L_\alpha\,L_\beta}_{\alpha,\,\beta}(k,k';E)=\sum_{b}g^{L_\beta\,*}_{\beta\,b}(k')\frac{1}{E-m_{b}}g^{L_\alpha}_{\alpha\,b}(k)+V^{J\,L_\alpha\,L_\beta}_{\alpha,\,\beta}(k,k')\,.
\end{equation}
The bound states or resonances are obtained by searching for the
poles of the $T$-matrix in the complex plane.

\section{Predictions of $0^+$ and $1^+$ $B_s$ states}
\label{sec:results}
In the extended HEFT framework, there are four undetermined
parameters: the $\gamma$ and the cutoff parameter $\Lambda^{\prime}$
in the QPC model, the coupling constant $g_c$ which combines the
$B^{(*)}B^{(*)}V$ and $KKV$ vertices, as well as the cutoff
$\Lambda$ from the $B^{(*)}\bar K$ interactions.
In our previous work~\cite{Yang:2021tvc}, we constructed the
Hamiltonians for the charm-strange mesons with $J^P=$ $0^+$ and
$1^+$ to simultaneously fit two sets of lattice data from
Refs.~\cite{Lang:2014yfa,Bali:2017pdv}.
When the cutoff $\Lambda$ was taken as 1 GeV, the other parameters
were fitted as
\be g_c = 4.2^{+2.2}_{-3.1}\;,\;\;\; \Lambda^{\prime} =
0.323^{+0.033}_{-0.031}\;\text{GeV}\;,\;\;\; \gamma =
10.3^{+1.1}_{-1.0}\,. \ee
These values are consistent with those in other phenomenological
investigations~\cite{Godfrey:2015dia,Shen:2019evi}.
For the $\Lambda$ dependence, it can be absorbed by the
renormalization of the interaction kernel.
With different $\Lambda$ employed, the final results remained the
same as shown in the supplemental material of
Ref.~\cite{Yang:2021tvc}.

\begin{figure*}[!htp]
\centering
\includegraphics[width=0.9\linewidth]{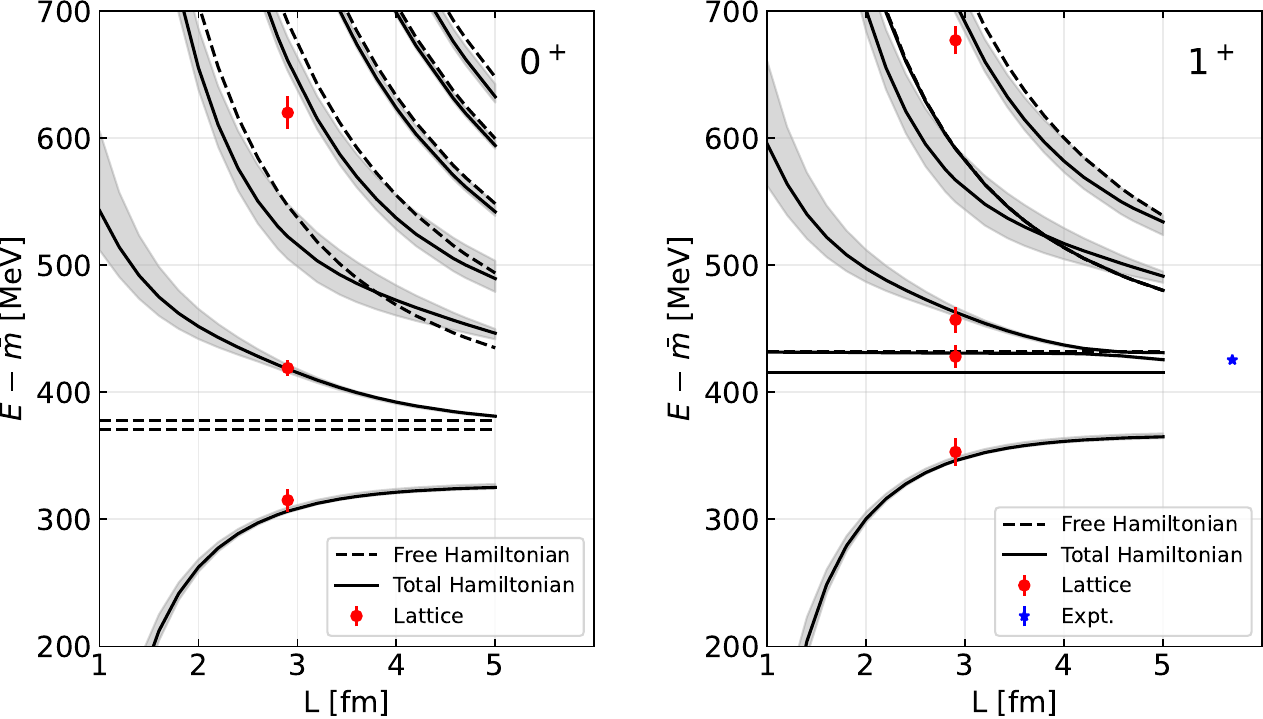}
\caption{
The comparison of the predicted energy levels for the $B^*_{s0}$
(left), the $B^{*(\prime)}_{s1}$ (right) states with lattice
simulation. The $\bar m$ is defined as  $\frac{1}{4}\left(m_{B_{s}}+3m_{B_s^{*}}\right)=5403.3$ MeV.
The red dots with the error bar are the lattice energy levels from
Ref.~\cite{Lang:2015hza}, while the blue star is the experimental
mass of $B^{*\prime}_{s1}$.
The black curves and the dashed lines are the predictions in
a finite volume with and without interacting Hamiltonian,
respectively. } \label{fig:spec}
\end{figure*}
%
With the heavy quark symmetry, we take the same parameters for the
bottom-strange mesons as those in the charm-strange sector.
The predicted energy levels of $J^P=0^+$ (left) and $1^+$ (right)
bottom-strange states are presented in Fig.~\ref{fig:spec}.
For comparison, we also present the lattice energy levels from
Ref.~\cite{Lang:2015hza} .
As shown in Fig.~\ref{fig:spec}, our predictions are well consistent
with the lattice simulation.
%

\begin{table*}
 \renewcommand\arraystretch{1.5}
\label{tab:mass}
 \centering
\begin{tabular}{c|ccc}
\hline\hline
$J^P$                        &             &          $0^+$       &       $1^+$                                  \\
\hline 
\multirow{7}{*}{mass [MeV]}
& rel. quark model~\cite{DiPierro:2001dwf}                     &                                $5804$                       &                                $5842$                                  \\
& rel. quark model~\cite{Ebert:2009ua}                     &                                $5833$                       &                                $5865$                                  \\
& rel. quark model~\cite{Sun:2014wea}                     &                                $5830$                       &                                $5858$                                  \\
&  nonrel. quark model~\cite{li:2021hss}                                           &                                $5788$                                 &                                $5810$                                  \\ 
&  quark model (KKMT)~\cite{Dmitrasinovic:2012zz}       & $5719$  & $5765$ \\
\\

& LO $\chi-SU(3)$~\cite{Kolomeitsev:2003ac}                     &                                $5643$                       &                                $5690$                                  \\
& Bardeen, Eichten, Hill~\cite{Bardeen:2003kt}                     &                                $5718\pm35$                       &                                $5765\pm35$                                  \\
& LO UChPT~\cite{Guo:2006fu,Guo:2006rp}                     &                                $5725\pm39$                       &                                $5778\pm7$                                  \\
& NLO UHMChPT~\cite{Cleven:2010aw}                          &                                $5696\pm20\pm30$              &                                $5742\pm20\pm30$                                  \\
& NLO UHMChPT~\cite{Du:2017zvv}                          &                                $5720_{-23}^{+16}$              &                                $5772_{-21}^{+15}$                             \\
& HQET + ChPT~\cite{Colangelo:2012xi}                    &                   $5706.6\pm1.2$              &                                $5765.6\pm1.2$                                  \\
& Covariant ChPT~\cite{Altenbuchinger:2013vwa}                &                                $5726\pm28$              &                                $5778\pm26$                                  \\

& local hidden gauge~\cite{Sun:2018zqs}                                      &                                $5475.4\sim5457.5$             &                                $5671.2\sim5663.6$                                  \\
& heavy meson chiral unitary~\cite{Albaladejo:2016ztm}                           &                                $5709\pm8$                         &                                $5755\pm8$                                  \\ \\
& lattice QCD~\cite{Gregory:2010gm}                     &                                $5752\pm16\pm5\pm25$                       &                                $5806\pm15\pm5\pm25$                                  \\
& lattice QCD~\cite{Lang:2015hza}                                 &                                $5713\pm11\pm19$                       &                                $5750\pm17\pm19$                                  \\ \\
& this work                                                              &                                $5730.2_{-1.5}^{+2.4}$        &                                $5769.6_{-1.6}^{+2.4}$                                  \\
\hline $P(\bar{b}s)[\%]$
& heavy meson chiral unitary~\cite{Albaladejo:2016ztm}     &                                $48.2\pm1.5/54.2\pm1.1$        &                                    $50.3\pm1.4/51.7\pm1.3$            \\
& this work                                          &                                $54.7_{-4.1}^{+5.2}$        &                                    $56.7_{-3.7}^{+4.6}$             \\
\hline\hline
\end{tabular}
\caption{
The comparison of the $B_s$ pole masses (MeV) and the contents of
bare cores extracted in this work with those from other theoretical
works and lattice QCD.
In this work, the content of the bare $\bar b s$ cores in the $B_s$
states, denoted as $P(\bar b {s})$, is extracted at $L=5$ fm. The errors on our masses and probabilities are obtained from the errors of the parameters in Eq.(3.1).}
\end{table*}

\begin{figure*}[!htp]
\centering
\includegraphics[width=0.5\linewidth]{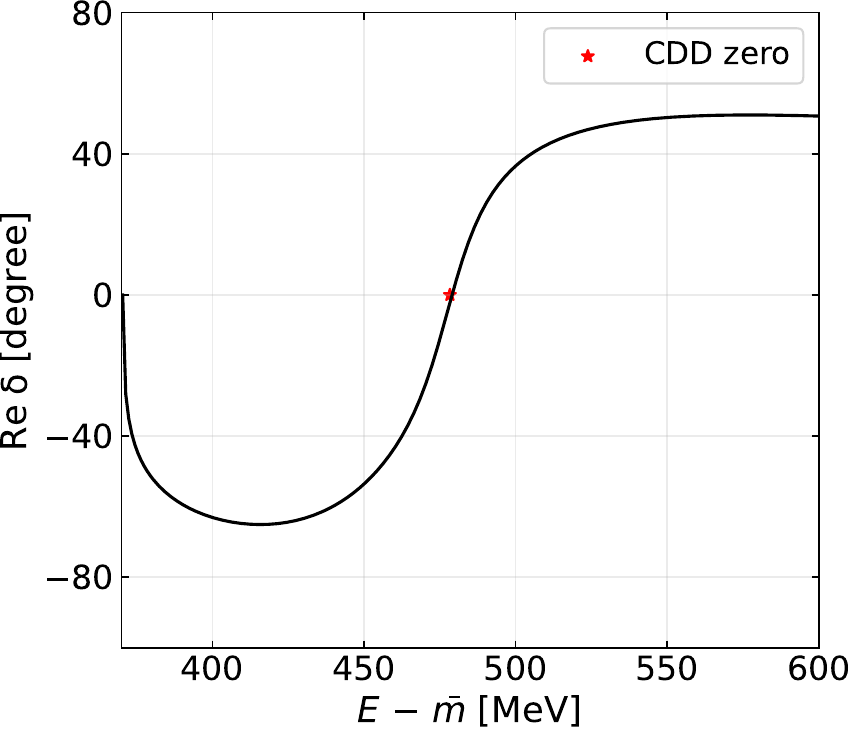}
\caption{ The phase shift of $B\bar K$ scattering, in which the red star
denotes the possible CDD zero. The $\bar m$ is defined as  $\frac{1}{4}\left(m_{B_{s}}+3m_{B_s^{*}}\right)=5403.3$ MeV.} \label{fig:phase}
\end{figure*}

In Fig.~\ref{fig:spec} (left), there is a special crossing of energy
levels (marked as $E_c$) between the free (dashed lines) and total
Hamiltonians (solid lines) in the $0^+$ sector around $L=3.7$ fm.
In principle, the free energy levels in the finite volume correspond
to the phase shifts of the scattering being $0$ or $\pi$ in the
infinite volume, i.e. there is no interaction at all.
Now the energy levels with and without the interaction of Hamiltonian share the same eigen-energy, which means that the system feels no
interaction at $E_c$ and correspondingly, we should have $T(E_c)=0$.
This is confirmed as shown in Fig.~\ref{fig:phase}, where we present
the dependence of phase shift on the center of mass energy and the
red star shows $\delta(E_c) = 0$.
Such energy $E_c$ is known as the Castillejo-Dalitz-Dyson (CDD)
zero~\cite{Castillejo:1955ed}.

The appearance of the CDD zero, which is strong evidence of the cancellation in the potential,  is promising for us to understand the
physical picture of the bound state or resonance. However, the details of such cancellations depend on the parameterization of the potential. 
In our framework, the contributions from the  coupling of
the bare state with the threshold channels and the channel-channel
potentials canceled at the CDD zero.
Hence, it indicates the existence of the bare state and also the important role of threshold channel components as discussed in
various
references~\cite{Baru:2010ww,Hanhart:2011jz,Guo:2016wpy,Kamiya:2016oao,Kang:2016jxw,Hyodo:2008xr,Hyodo:2013nka}.
However, until now, there is no exact and convincing evidence for
the existence of CDD zero. The main reason is that the scattering $T$-matrix of $2\rightarrow 2$ progress in the heavy quark sector cannot be obtained directly in experiments. 
Here, we provide a novel method to search for a CDD zero in lattice QCD, checking the
the crossing point of the energy levels with and without interaction
Hamiltonian in the finite volume.
As shown in our model, the spectrum of the $B^*_{s0}$ state provide
a golden platform to confirm the existence of the CDD zero.

In Ref.~\cite{Albaladejo:2016ztm}, the authors also considered the bare $\bar b s$ core and the $B^{(*)}\bar K$ component with a different potential parametrization,  where two sets of bare masses were extracted from other quark model calculations and the other two referred parameters were determined by fitting six data points from lattice QCD~\cite{Lang:2015hza}. In this work, we directly employ the parameters determined in the $D_s$ sector to study the $B_s$ sector with the heavy quark flavor symmetry.  The predicted energy levels are surprisedly consistent with both the $0^+$ and $1^+$ lattice QCD energy levels at low energy. With the two different potential parameterizations, the mass spectra and components of the two $B_s$ states are similar to each other as summarized in Table~\ref{tab:mass}. However, the two parameterizations provide different results for the CDD pole.  Our model predicts the existence of a CDD zero in the $0^+$ sector, while it is absent in Ref.~\cite{Albaladejo:2016ztm}. The third lattice data in the $0^+$ sector corresponds to the third energy level in  Ref.~\cite{Albaladejo:2016ztm}, while it is close to the fourth one in our work. As described in the lattice paper \cite{Lang:2015hza}, three $S$-wave meson-meson operators were included with two different momentum sets, $\vec p=0$ and $\vec{p}=\pm (2\pi/L) \hat{e}_{x,y,z}$, labeled by $B(0)\bar{K}(0)$ and $B(1)\bar{K}(-1)$, respectively.
The lowest energy level is dominated by several quark-antiquark operators and energy level 2 by the one of $B(0)\bar{K}(0)$ operator and another quark-antiquark operator. The highest energy level is purely dominated by the $B(1)\bar{K}(-1)$ operator.
The energy shift between the lattice level and the corresponding $B\bar{K}$ free energy is due to the coupled channel effect, which should be smaller with increasing $p$ of $B\bar{K}$. 
However, the shift between the third lattice energy level and the $B(1)\bar{K}(-1)$ free energy is even larger than that between the second lattice energy level and the $B(0)\bar{K}(0)$ free energy.
Therefore, it might miss another energy level between the highest two lattice data.
More lattice QCD calculations are expected in the future to give more constraint to the parameterizations and examine the CDD pole's existence.

At last, we obtain the pole masses of the $T$-matrix and list them
in Table~\ref{tab:mass}, together with the masses from other
phenomenological studies and lattice QCD calculation for comparison.
The pole positions of $B^*_{s0}$ and $B^*_{s1}$ are located in the
first Riemann-sheet at $5730.2_{-1.5}^{+2.4}$ MeV of the $B\bar K$ channel and $5769.6_{-1.6}^{+2.4}$
MeV of the $B^*\bar K$ channel, respectively.
Both the $0^+$ and the lighter $1^+$ bare $\bar b s$ core have a
significant mass shift due to the $S$-wave interactions with
$B^{(*)}\bar K$ channel.
The coupled channel effects also make them a mixture of the bare
$\bar b s$ core and $B^{(*)}\bar K$ components.
By analyzing the eigenvectors, the $P(\alpha)$ shows that the two
components are significant and essential for the $B_{s0}^*$ and
$B_{s1}^*$ states.
The bare $\bar b s$ core in the $B^*_{s0}$ accounts for around
$54.7_{-4.1}^{+5.2}\%$ at $L=5$ fm, while the $B\bar K$ component occupies around
$45.3\%$.
The bare $\bar b s$ core in the $B^*_{s1}$ accounts for around
$56.7_{-3.7}^{+4.6}\%$ at $L=5$ fm, while the $B^*\bar K$ component occupies around
$43.3\%$.

In contrast, the $D$-wave interaction around the threshold is
significantly suppressed at $\mathcal O(k^2)$ compared with the
$S$-wave one.
Therefore, the energy level of the $B_{s1}^{*\prime}$ almost keeps
stable, and its bare $\bar b s$ core dominates.

\section{Summary} \label{sec:summary}

In summary, we have investigated the $0^+$ and $1^+$ bottom-strange
mesons with the framework which incorporates the quark model, the
QPC model, and the coupled-channel unitary approach into the HEFT
framework.
This framework has been successfully used to describe both the
lattice QCD data and the experimental mass spectra of the
$D^*_{s0}(2317)$, $D^*_{s1}(2460)$,  $D^*_{s1}(2536)$, and
$D^*_{s2}(2573)$ states.
Here, we employed the same parameters determined by fitting the
lattice energy levels of the $D_s$ states.
The predicted energy levels of the $0^+$ and $1^+$ $B_s$ states are
well consistent with the lattice QCD simulation at low energy.
Moreover, a very clear physical picture emerges from our results for
the $0^+$ and $1^+$ $B_s$ states, i.e., they are the mixture of the
bare $\bar b s$ and $B^{(*)}\bar K$ components.
The bare masses are shifted by tens of MeV due to the
coupled-channel effects with the $S$-wave $B\bar K$ and $B^{*}\bar K$
channels, respectively.

The extracted pole masses from the $T$-matrix are also consistent
with the results from the lattice QCD and other phenomenological
models as shown in Table~\ref{tab:mass}.
In addition, we predict a CDD zero in the $B\bar K$ scattering through
the finite volume spectrum.
It can be used to examine the potential as well as the inner
structures of  the physical states  as pointed out in
Ref.~\cite{Li:2022aru}.
Therefore, future investigations from a theoretically motivated
model and lattice QCD simulation for the $B\bar K \to B\bar K$ process is
necessary and expected.

\begin{acknowledgments}

We thank the useful discussions and valuable comments from  Yan Li and Mao-jun Yan. 
This work is partly supported by the National Natural Science
Foundation of China (NSFC) under Grants Nos. 12275046 (Z.Y.),
and by the Natural Science Foundation of Sichuan Province under Grant No. 2022NSFSC1795 (Z.Y.),
and the supported by JSPS KAKENHI under Grant No.
20F20026(G.J.W.),
and by the National Natural Science Foundation of China under Grants Nos.
12175239 and 12221005(J.J.W.),
and by the National Key R$\&$D Program of China under Contract No.
2020YFA0406400 (J.J.W.),
and by the JSPS KAKENHI under Grant Nos.~19H05159, 20K03959, and
21H00132 (M.O.),
and by the National Natural Science Foundation of China under Grant
Nos.11975033 and 12070131001 (S.L.Z.).

\end{acknowledgments}

\bibliography{Bs.bib}
\end{document}